# Two-dimensional Topological Semimetals Protected by Symmorphic Symmetries

Wei Luo[1,2†], Junyi Ji[1,2†], Jinlian Lu[1,2], Xiuwen Zhang[3*], and Hongjun Xiang[1,2*]

[1]*Key Laboratory of Computational Physical Sciences (Ministry of Education), State Key Laboratory of Surface Physics, and Department of Physics, Fudan University, Shanghai 200433, P. R. China*

[2]*Collaborative Innovation Center of Advanced Microstructures, Nanjing 210093, P. R. China*

[3]*College of Electronic Science and Technology, Shenzhen University, Shenzhen 518060, P.R. China*

**Email:** hxiang@fudan.edu.cn, xiuwenzhang@szu.edu.cn

[†]W.L. and [†]J. J. contributed equally to this work.

**Abstract**

Two-dimensional (2D) band crossing semimetals (BCSMs) could be used to build a range of novel nanoscale devices such as superlenses and transistors. We find that symmorphic symmetry can protect a new type of robust 2D BCSMs, unlike the previously proposed 2D essential BCSMs protected by non-symmorphic symmetry [Young *et al.*, Phys. Rev. Lett. 115, 126803 (2015)]. This type of symmorphic symmetry protected (SSP) 2D essential BCSMs cannot be pair annihilated without destroying the crystalline symmetries, as opposed to the 2D BCSMs caused by the accidental band crossing. Through group theory analysis, we find that 2D SSP BCSMs can only exist at the K (K') point of Brillouin zone (BZ) of four layer groups and identify nonmagnetic 2D $FeB_2$ as a candidate. Interestingly, nonmagnetic 2D SSP BCSMs can host a single pair of band crossing points (BCPs), whereas nonmagnetic three-dimensional (3D) Weyl semimetals (WSMs) have at least two pairs of band crossing Weyl points. It is found that the single pair of BCPs are robust against any kinds of strain. Furthermore, our calculation suggests that essential 2D SSP BCSMs can be used to realize electric field control of spin-texture, thus are promising candidates for spintronic devices.

## I. INTRODUCTION

Three-dimensional (3D) Weyl semimetals (WSMs) [1, 2], in which two bands cross with each other linearly in momentum space near the Fermi level, have attracted great interest due to their novel transport properties, such as chiral anomaly effect [3-5], anomalous Hall effect [6-10], and intrinsic spin Hall effect [11]. More interestingly, 3D WSMs host the topologically protected Fermi-arc [1, 6, 12], which are distinguished from the topological surface states of topological insulators and Chern insulators. Weyl points (WPs) in 3D WSMs must exist in pairs due to the conservation of chirality [13-15] which is forced by the no-go theorem [16], hence are robust to perturbation without breaking the translational symmetry [17]. The only way to annihilate them is letting them meet with each other at a point in the Brillouin zone (BZ), consequently, forming a Dirac point or opening a band gap.

Although 3D WSMs were widely studied in recent years [2, 8, 17-24], there are only a few studies on their analogies [25-29], i.e. “2D band crossing semimetals (BCSMs)”. Similar with the 3D WPs, the 2D band crossing WPs also maintain linear dispersions near band crossing points (BCPs). 2D BCSMs could display extraordinary properties, for instance, the nonlinear anomalous Hall effect in 2D $WTe_2$ [30]. Generally speaking, considering whether BCPs can be pairwise annihilated or not without breaking the system’s crystalline symmetries, 2D BCSMs can be classified into two categories. For the first category, the BCPs are caused by accidental band crossing and can be pairwise annihilated while preserving the crystalline symmetries. Park and Yang systematically investigated this kind of 2D BCSMs [31]. The potassium doped few-layer black phosphorus [32], 1*T’* monolayer of $WTe_2$ with an electric field perpendicular to the 2D plane [30] and 2D $TiB_2$ [33] belong to this category. For the second category, the BCPs cannot be pairwise annihilated as long as the crystalline symmetries are preserved—we thus refer them as 2D essential BCSMs. This kind of BCSMs have been theoretically discussed in non-symmorphic layer groups (LGs) [34]. However, to the best of our knowledge, no material candidates have been proposed for the nonsymmorphic symmetry-protected 2D essential BCSMs. For symmorphic LGs, it was widely believed that the 2D essential BCSMs can only be enforced by the Kramers theorem [34, 35].

In this article, we propose for the first time that 2D essential BCSMs can emerge in symmorphic LGs, which arise from the spatial symmetries instead of the Kramers degeneracy. Based on systematic group theory analysis, we find that SSP 2D essential BCSMs can only exist at the K (K') point of the BZ of four LGs (67, 70, 76 and 77). Interestingly, 2D SSP essential BCSMs can host a single pair of BCPs when preserve the time-reversal (TR) symmetry and break the inversion symmetry, different from the 3D time-reversal invariant (TRI) WSMs that host at least two pairs of WPs [36]. This character makes the SSP 2D essential BCSMs an ideal platform for detecting the BCPs. We note that the 2D $FeB_2$ material [37] (layer group index 77) hosts only one pair of ideal BCPs at the Fermi level which are protected by the pure symmorphic symmetries without involving the TR symmetry. Interestingly, we find that the BCPs in $FeB_2$ are robust for large uniaxial strain (~25%) which is similar with the robustness of Dirac points in graphene under strain [38, 39].

## II. Design principles for the SSP 2D essential BCSM from group theory analysis

BCSMs are formed due to the doubly degenerate band crossing points in BZ of materials with either broken TR or spatial inversion symmetry (Similar to that of 3D WSMs). For symmorphic LGs, there are no non-symmorphic symmetries to stick a certain minimal number of bands together [34, 40]. In this case, it is believed [40] that only TR symmetry can group two bands together (filling of odd number of electrons will lead to WPs at TRI points). However, if we inspect the character table of irreducible representations (IRs) for symmorphic LGs, we find that there exist some two-dimensional IRs. This means that 2D essential BCPs may be maintained solely by symmorphic symmetries. Note that two-dimensional IRs for symmorphic LGs are only available at high symmetry points and lines. However, the 2D essential BCPs can only exist at high symmetry points instead of lines since the two-fold essential degeneracy must split along different directions near the BCPs. For 2D systems, the high symmetry points in BZ include only six points. They are Γ(0,0) X(0.5,0), Y(0,0.5), M(0.5,0.5),

K(1/3,1/3) and K'(-1/3,-1/3). Since we are focusing on the SSP 2D essential BCSMs, the BCPs located at the TR invariant points (Γ, X, Y, and M) that are protected by the TR symmetry[35] are excluded. Hence, the SSP essential BCPs must locate at K(1/3,1/3) or K'(-1/3,-1/3). Namely, the SSP 2D essential BCSMs can only exist in the hexagonal/trigonal systems. For the SSP 2D essential BCSMs, there are two different cases. For the first case, a band is singly degenerate near K (K') in the BZ without considering the spin-orbit coupling (SOC) effect. After considering the SOC effect, in order to realize the BCSMs phase, the IR of this band should belong to a two-dimensional IR at the K (K') point and decompose to two one-dimensional IR near K (K') [see Appendix A]. Note that the BCPs in this case have the same electron filling condition with that of TR protected BCPs (i.e. filling of odd number of electrons). Hence, we will not discuss this case hereafter (details about this kind of BCPs can be seen from Appendix A). For the second case, without considering the SOC effect, the band belongs to a two-dimensional IR at K (K'). After considering the SOC effect, the two-dimensional IR will decompose to two two-dimensional (2+2 mode, Fig. 1a) or one two-dimensional and two one-dimensional (2+1+1 mode, Fig. 1b) IRs according to their symmetry at the K (K') point. For the "2+2" mode, it will cause a band gap at the K point, so we exclude this case. For the "2+1+1" mode (the band decomposition can be deduced based on the group theory, see Appendix A), due to the different SOC form, the energy levels for the one doubly degenerated as well as the two singly degenerated bands can be different. Here, we are only interested in the case where the BCPs locate at the Fermi level as illustrated in Fig. 1b. It is noted that our discussion on SSP 2D essential BCSMs is limited to TRI systems. Based on the above analysis, for realizing the SSP 2D essential BCSMs, the system should satisfy four conditions: (I) the breaking of inversion symmetry since the system adopts TR symmetry; (II) The lattice maintains the hexagonal or trigonal symmetry; (III) The little group of the K (K') point should have a two-dimensional IR without considering SOC effect, which decomposes to one two-dimensional and two one-dimensional IR after considering SOC; (IV) The band dispersion near the crossing point (i.e., K or K') should be linear. The symmorphic LGs can be obtained by removing the T and O point groups from the 3D point groups. From the table of 80-LGs [41], one can easily obtain the 36 symmorphic LGs. Considering first condition, i.e., removing the groups with inversion symmetry, there remain 25 LGs. Then, the second condition requires that the lattice maintains the hexagonal/trigonal symmetry (with $C_3$ or $C_6$ rotation axis). Thus, there remain 11 LGs. Considering the third condition, the number of candidate LGs shrinks to 4. Their indexes are 67, 70, 76 and 77, corresponding to $D_3$-2, $C_{3v}$-2, $D_6$ and $C_{6v}$ point groups. For the last condition, if the symmetric Kronecker product $[R_k \times R_k]$ ($R_k$ represents the two-dimensional IR at K (K') in the BZ ) of the two-dimensional IR with itself contains the vector representation of $G_k$ [$G_k$ is the little group at the K (K') point], the linear dispersion can be guaranteed [42]. We find that this condition does not exclude any of the remaining four LGs (67, 70, 76 and 77). Based on the above analysis, the searching of SSP 2D essential BCSMs can be conducted as follows (see Fig. 1b): first, finding the band crossing belonging to a two-dimensional IR at K without considering the SOC effect. Then, finding the cases in which after including the SOC effect, the two-dimensional IR splits into a two-dimensional IR with linear dispersion and two one-dimensional IRs as illustrated in Fig. 1b.

## III. 2D essential BCPs in the FeB2 system

We find that the $FeB_2$ monolayer [37] (see Fig. 2b) with graphene-like boron sheet is a candidate SSP 2D essential BCSMs which belongs to the No. 77 ($C_{6v}$) layer group. The band structures of $FeB_2$ calculated by using first-principles calculations (see Appendix B) without/with SOC effect are shown in Fig. 2c and 2d, respectively. From Fig. 2c, we can see that the wave function belongs to the two-dimensional IR $\Gamma_3$ of the single group of $C_{3v}$ at the $K(1/3,1/3)$ point, and that the band crossing is mainly contributed by the Fe $(d_{x^2-y^2}, d_{xy})$ and $(d_{xz}, d_{yz})$ orbitals. Here, it is noted that the $(d_{x^2-y^2}, d_{xy})$ and $(d_{xz}, d_{yz})$ orbitals maintain the same transformation character under the $C_{3v}$ point group. In other words, they belong to the same two-dimensional IR $\Gamma_3$ of $C_{3v}$. After considering the SOC effect, the direct product of $\Gamma_3$ and spin representation $D_{1/2}$ is decomposed into one $\Gamma_4$ (one-dimensional) and one $\Gamma_5$ (one-dimensional), and one $\Gamma_6$ (two-dimensional) IRs (see Fig. 2d). Since the energy order is computed to be $\Gamma_4 < \Gamma_6 < \Gamma_5$, the Weyl point locates exactly at the Fermi level, as shown in Fig. 2d, forming a SSP 2D ideal essential BCSM.

## IV. Low-energy effective Hamiltonian of FeB2

At the *K* point, the little point group is $C_{3v}$ and the crossing bands belong to the $\Gamma_3$ (two-dimensional) IR without considering the SOC effect. Based on the theory of invariants [43], we derive the low-energy effective Hamiltonian of 2D $FeB_2$ system with/without SOC effect. Without SOC effect, the bases which transform according to the $\Gamma_3$ IR can be written as $|\phi_1\rangle$ and $|\phi_2\rangle$. The low-energy effective Hamiltonian at the *K* point is written as:

$$H_K = \begin{pmatrix} C_0 & C_2 k_+ \\ C_2 k_- & C_0 \end{pmatrix}$$

Where $k = k' - K$ and $k'$ is a point near the *K* point ($k_\pm = k_x \pm i k_y$). $C_0$ and $C_2$ are constants. It is clear that the two bands with the eigenvalues $E_{1,2}(k) = C_0 \pm C_2\sqrt{k_x^2 + k_y^2}$ degenerate with each other at the *K* point. From the eigenvalues, one can see that the two bands maintain isotropic linearly dispersion near the *K* point. After considering the SOC effect, since the spin degree is included, the bases change to be $|\phi_1 \uparrow\rangle$, $|\phi_1 \downarrow\rangle$, $|\phi_2 \uparrow\rangle$ and $|\phi_2 \downarrow\rangle$. The low-energy effective Hamiltonian (details can be seen from Appendix C) at the $K(1/3,1/3)$ point becomes a $4 \times 4$ matrix:

$$H = \begin{pmatrix} c_0 & -\frac{a_1+b_1}{\sqrt{2}} k_- & -i\frac{a_1-b_1}{\sqrt{2}} k_- & i c_1 k_+ \\ -\frac{a_1+b_1}{\sqrt{2}} k_+ & \frac{a_0+b_0}{2} & i\frac{a_0-b_0}{2} & i\frac{a_1-b_1}{\sqrt{2}} k_- \\ i\frac{a_1-b_1}{\sqrt{2}} k_+ & -i\frac{a_0-b_0}{2} & \frac{a_0+b_0}{2} & \frac{a_1+b_1}{\sqrt{2}} k_- \\ -i c_1 k_- & -i\frac{a_1-b_1}{\sqrt{2}} k_+ & \frac{a_1+b_1}{\sqrt{2}} k_+ & c_0 \end{pmatrix},$$

where $a_0, a_1, b_0, b_1, c_0$ and $c_1$ are real constants, and $\mathrm{i} = \sqrt{-1}$. One can see that the eigenvalues are $\mathrm{E}_{1,2} = \mathrm{c}_0$, $\mathrm{E}_3 = \mathrm{a}_0$ and $\mathrm{E}_3 = \mathrm{b}_0$ at the $K(1/3,1/3)$ point. $\mathrm{E}_{1,2}$ is two-fold degenerated, whereas $E_3$ and $E_4$ are not degenerated with each other at $K(1/3,1/3)$ and there exists a $|b_0 - a_0|$ energy gap between them. Away from the $K$ point, the dispersion of the two bands close to the Fermi level is linear (see Appendix A and C). From the effective Hamiltonian analysis, we can easily see that, due to the SOC effect, the spinless BCPs (not including spin) changes into the spinful BCPs (including spin), which agrees with the motif described by Fig. 1b.

## V. The emergence of a single pair of BCPs in 2D non-magnetic systems

For these four-layer groups, the shape of their BZ are the same, as shown in Fig. 2a. Since the K and K' are related by the TR symmetry, BCPs must locate at K and K' simultaneously, leading to a single pair of BCPs that is very different from the 3D WPs. For 3D non-magnetic WSMs, there exist at least two pairs of WPs in the BZ [36], because a Weyl point located at a k point in the BZ is converted into another Weyl point at the –k point with the same chirality due to the TR symmetry. Based on the no-go theorem [13], there must exist another pair of WPs with the opposite chirality. However, for 2D BCPs, the concept of the chirality of a "magnetic monopole" is not well defined. Hence, a single pair of BCPs in a 2D non-magnetic system is allowed. This new feature of 2D BCSMs provides a good platform for experimental study of the 2D BCPs. It is noted that the number of 2D BCPs must be even since the sum of Berry phases around BCPs must equal to zero (or integral multiple of $2\pi$) [25] in the whole 2D BZ. In addition, by combining the maximally-localized Wannier functions (MLWF) [44, 45] and wanniertools [46], we find that there are non-trivial edge states for zigzag $FeB_2$ nanoribbons (see Appendix D).

## VI. Strain-robust 2D BCPs in FeB2 system

We checked the stability of BCPs under different kinds of strain and find that they are robust against any kinds of strain (uniaxial and shear strain). Especially, BCPs in 2D $FeB_2$ cannot be annihilated even under 25% uniaxial tensile (reducing $C_{6v}$ to $C_2$) strain. For comparison, we plot band structures near the BCPs without/with 5% tensile strain with considering SOC effects (Fig. 3). One can easily find that BCPs have shifted away from the high symmetry K point under strain. Another interesting result is that one pair of BCPs (without strain) evolve into two pair of BCPs under stain. As shown in Fig. 3b, two BCPs are very close to each other under strain. Note that there is only one pair of BCPs under strain if we do not consider the SOC effect. The formation of two pair of BCPs with considering SOC effects is forced by the rotation anomaly [47]. The robustness of the BCPs in the $FeB_2$ system under strain without considering SOC effects is similar with that of the robust Dirac points in graphene [38] which is protected by the space-time inversion symmetry [48-50]. Here, we find that BCPs in $FeB_2$ is protected by the combined operation of $C_{2z}$ rotation and TR symmetry. We check the different kinds of strain and find that they do not destroy the $C_{2z}$ rotation symmetry, thus, the $C_{2z}T$ combined symmetry is still maintained, leading to the robustness of BCPs under different kinds of strain. Since the 67 ($D_3$-2) and 70 ($C_{3v}$-2) LGs does not include a $C_{2z}$ rotation axis, the BCPs in these two groups are not robust under uniaxial and shear strain, i.e. for SSP 2D essential BCSMs, only the 76 and 77 LGs can maintain the robust BCPs under various strains. It is noted that the strain here changes the category of BCPs (i.e. from the second category to the first category). After adding uniaxial tensile strain, the BCPs are caused by accident band crossing and can be eliminated by merging two BCPs at the TR invariant points [39]. Note that here the accident band crossing is protected by the combined operation of $C_{2z}$ and TR symmetry that is different than the BCPs located at high symmetry line or plane, which is caused by two different one-dimensional IRs (the crossing bands maintain different eigenvalues) and is locally protected by a single crystal symmetry (such as rotation and mirror symmetry) without involving TR symmetry.

## VII. Spin-texture controlled by electric field

Since the $FeB_2$ monolayer adopts a buckled crystal structure, it has an intrinsic non-zero electric polarization along the *z* direction. Hence, we speculate that it may maintain a Rashba-like spin-texture. Using first-principles calculations, we calculate the spin-texture for 2D $FeB_2$ system (Fig. 4). One can see that the first valence band maximum $VBM_1$ [first conduction band minimum ($CBM_1$)] and second valence band maximum $VBM_2$ [second conduction band minimum ($CBM_2$)] adopt opposite spin-texture helicities. However, for $VBM_1$ and $CBM_1$, they maintain the same spin-texture helicity. As K and K' also adopt the same spin-texture helicity due to the TR symmetry, the electronic states near the Fermi surface of $FeB_2$ monolayer maintain a single spin-texture helicity. It is noted that here the spin expectation value for $VBM_2$ and $CBM_2$ should be zero at the K and K' points. This can be understood easily based on the group theory. The K and K' points have $C_{3v}$ little group. The invariant representation of $C_{3v}$ does not include the pseudo-spin vector (in another word, $C_{3v}$ is a non-pseudo-polar point group [51]). Hence, the spin should be zero for singly degenerated band $VBM_2$ and $CBM_2$. After reversing the electric polarization in the $z$ direction, i.e., switching the position of Fe and B atoms, the spin-texture are totally reversed, thus, realizing the electric field control of Rashba-like spin-texture. This can be understood as follows: The two states (polarization along z and –z, respectively) are related by an in-plane mirror symmetry operation, which does not change the k-point, but reverses the in-plane components of the spin vectors. This result can also be deduced from our previous low-energy effective Hamiltonian. We note that the band splitting away from the Weyl point is also fundamentally different from the usual Rashba splitting: (1) it does not take place at a TRI k-point; (2) the spin textures of the two bands near the Fermi level have the same chirality.

It is noted that although this system is metallic (strictly speaking, semimetallic), the electric polarization might be reversed since the electrons are not allowed to move along the $z$ direction [52, 53] and the electronic screening effect for the electric field is weak due to the semimetallic nature of the electronic structure. Hence, our works suggest the reversal of the spin-texture in a 2D BCSMs by an external electric field, which may broaden the way for topological quantum spintronics.

## VIII. CONCLUSIONS

In summary, we demonstrated that 2D SSP essential BCSMs can only exist in 67, 70, 76 and 77 layer groups. Interesting, this kind of 2D BCSMs can maintain a single pair of BCPs in the BZ. We find that the 2D buckled $FeB_2$ belonging to the layer group 77$^{th}$, is a candidate for 2D SSP essential BCSMs. Remarkably, the BCPs in this system are robust for different kinds of strain, and even the 25% uniaxial tensile strain cannot annihilate the BCPs. Moreover, our calculations indicate that the $FeB_2$ system adopts a Rashba-like spin-texture near the K (K') point and a single spin-texture helicity for all the electronic states near the Fermi surface. When the polarization is reversed, the Rashba-like spin-texture helicity is also reversed, realizing the electric field control of Rashba-like spin-texture in semi-metallic systems. Since the BCPs exactly locate at the inequivalent K and K' valleys, this kind of 2D essential BCSMs represent a good platform for studying topological valleytronics.

**ACKNOWLEDGMENTS**

Work at Fudan is supported by NSFC 11825403, the Program for Professor of Special Appointment (Eastern Scholar), the Qing Nian Ba Jian Program, and the Fok Ying Tung Education Foundation. Work at Shenzhen is supported by NSFC 11774239 and 61827815.

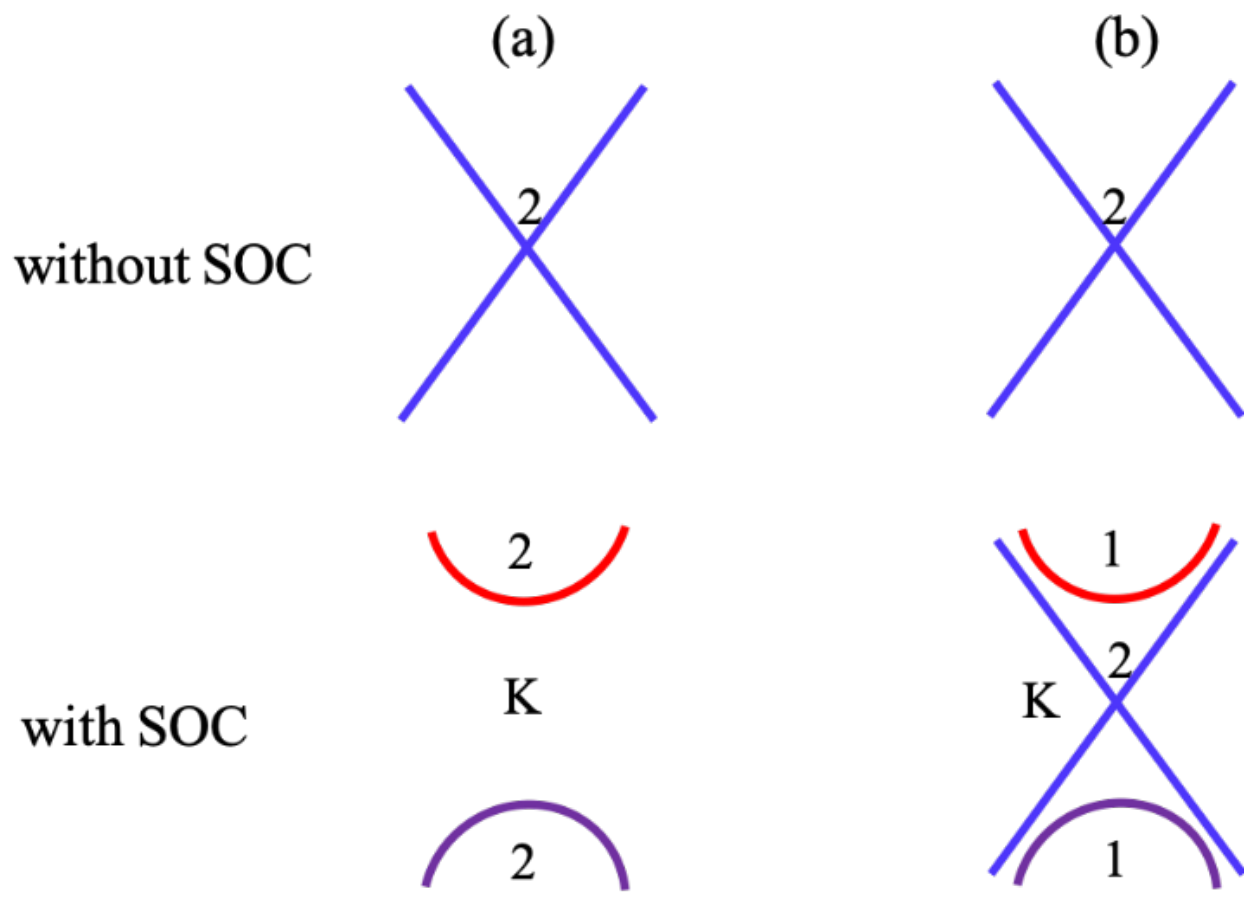

FIG. 1 Schematic diagram for searching the 2D SSP BCSMs. (a) Without SOC effect, the band is doubly degenerated at K point. After considering the SOC effect, the doubly

degenerated bands split into two two-fold degeneracy bands and form a band gap at K point. (b) After considering the SOC effect, the doubly degenerated bands split into one two-fold degeneracy band and two one-fold degeneracy bands. Thus, form the BCPs at the K point.

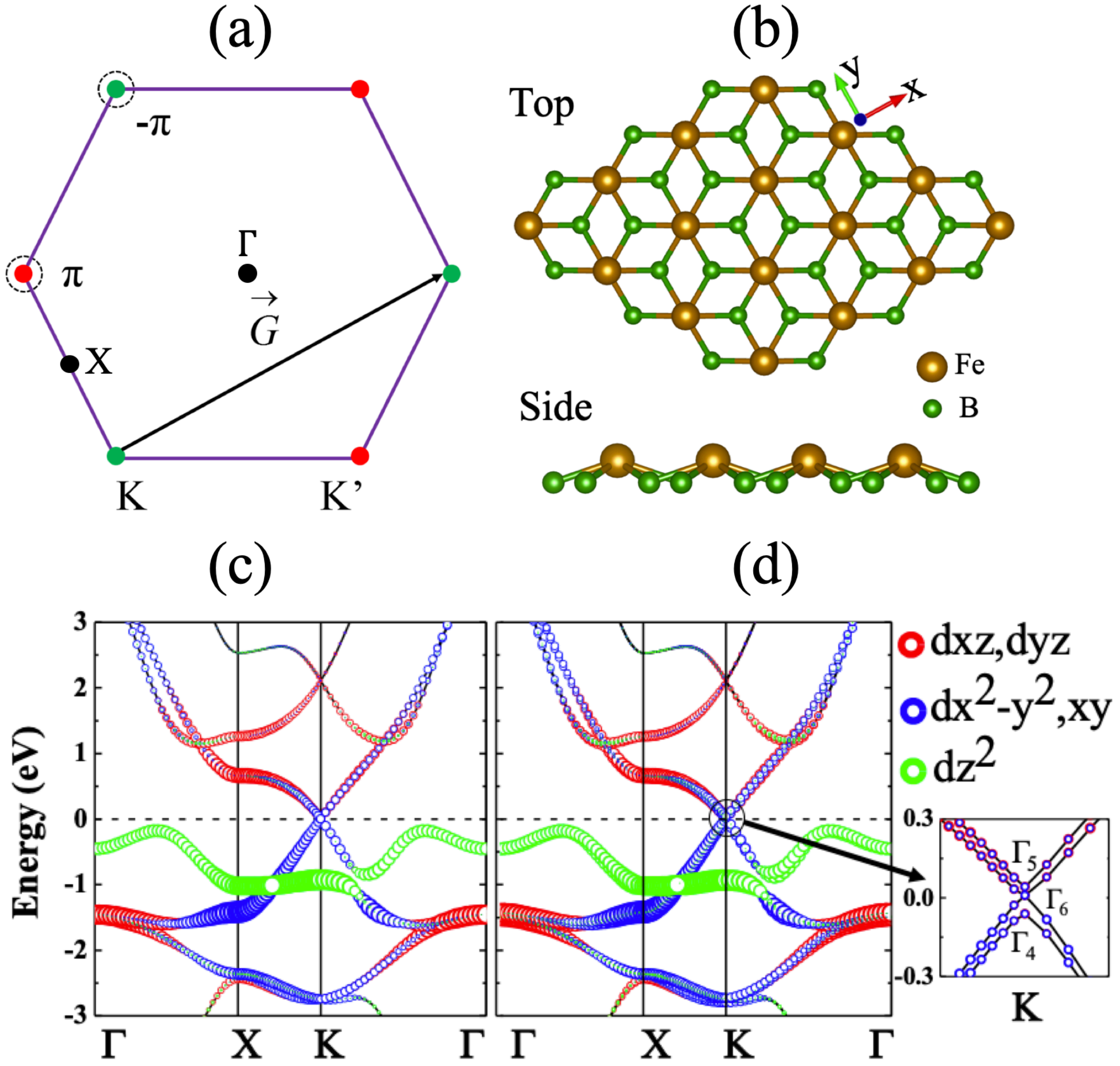

FIG. 2 (a) The BZ for 67, 70, 76 and 77 LGs. The two BCPs are located at the K and its time-reversal partner K'. (b) Top and side views of 2D $FeB_2$ system. (c) Band structure for the 2D $FeB_2$ system without considering the SOC effect. The crossing bands are contributed by the ($d_{xz}$, $d_{yz}$) and ($d_{x2-y2}$, $d_{xy}$) orbitals. (d) Band structure with the SOC effect taken into account. The bands split into two one-fold degenerated bands and one two-fold degenerated band, forming the ideal band crossing semimetal.

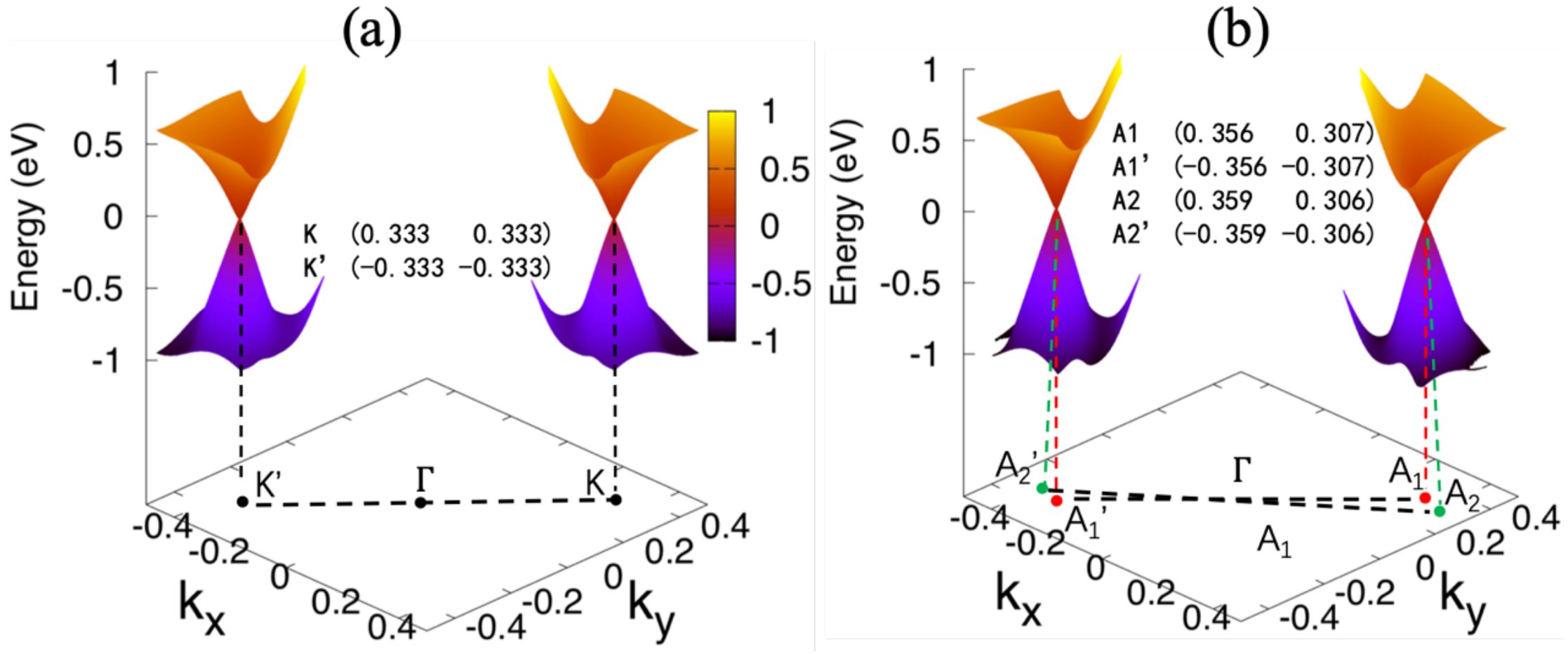

FIG. 3 The band structure for $FeB_2$ system near the K and K' points with SOC effects. (a) without strain and (b) with 5% uniaxial [along the x (zigzag) direction, see Fig. 2b] tensile strain. After adding the tensile strain, the BCPs shift their positions and evolve into two pair of BCPs.

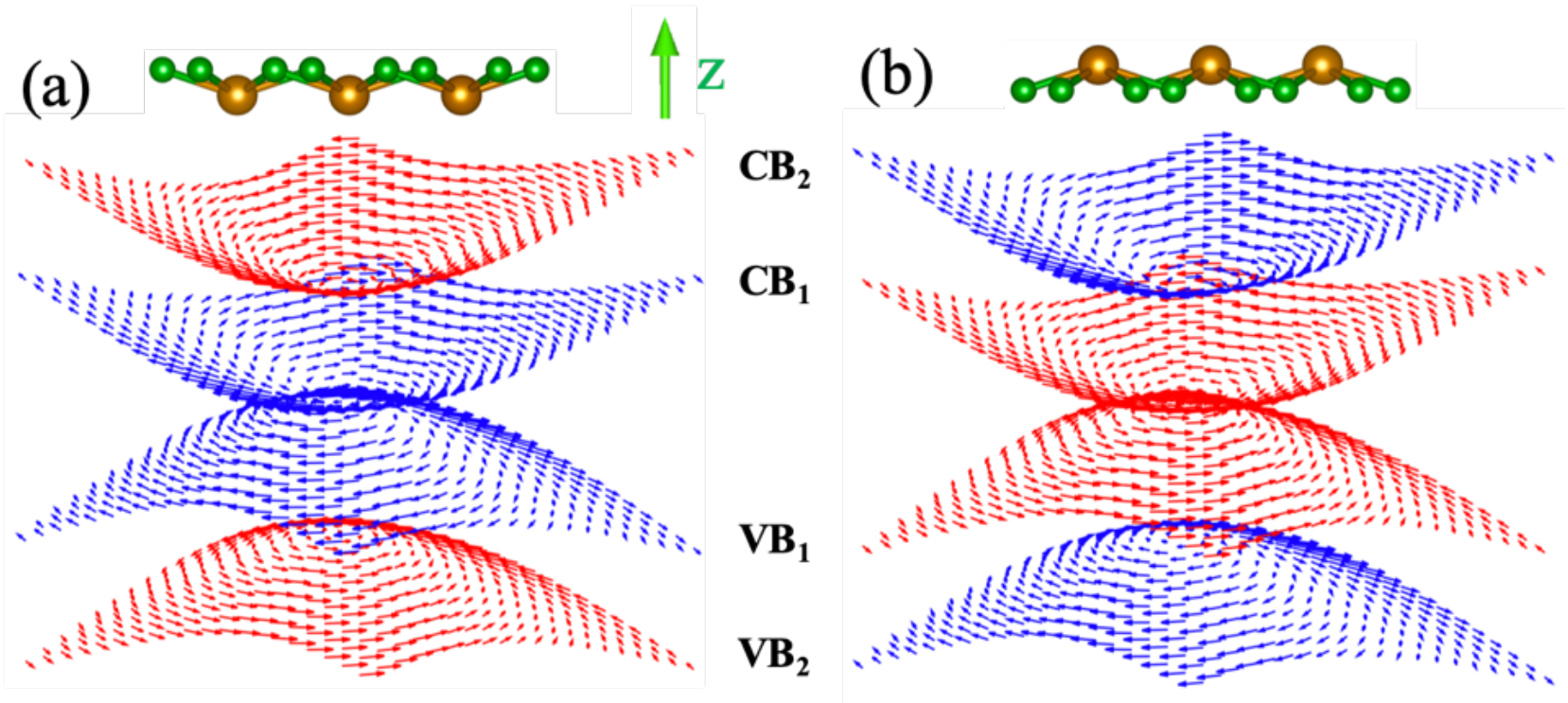

FIG. 4 The Rashba-like spin-texture of the top two valence band (VB) and lowest two conduction band (CB) for the $FeB_2$ system with polarization along the -z (a) and +z (b) direction. Note that the energy level of $VB_2$ and $CB_2$ have been shifted 0.1eV relative to that of $VB_1$ and $CB_1$ in order to make the spin-texture clearer. The blue and red color indicate clockwise and anti-clockwise, respectively. One can easily find the helicity of the spin-texture has been reversed due to the reverse of the polarization.

## APPENDIX A: Detailed group theory analysis on the condition of the existence of BCPs

As shown in the main text, four layer groups (Nos. 67, 70, 76, and 77) may host 2D Weyl points. Here, we now demonstrate in detail that this is indeed the case. For layer groups 67 and 76 ($D_3$-2 and $D_6$), the little group of K(K') point is $D_3$. For layer group 70 and 77 ($C_{3v}$-2 and $C_{6v}$), the little group of K(K') point is $C_{3v}$. Our analysis shows that the $D_3$ case is similar to the $C_{3v}$ case. Here, we take the $C_{3v}$ point group to illustrate the detailed analysis process. The character table of $C_{3v}$ is shown in Table I. The letter O represents the different classes for the $C_{3v}$ point group with considering the spin degree. $\Gamma_1$, $\Gamma_2$ and $\Gamma_3$ are the representation of single group of $C_{3v}$ point group. $\Gamma_4$, $\Gamma_5$ and $\Gamma_6$ are additional representations which involve the SOC effect. $D_{1/2}$ represents the spin representation.

First, we discuss the first case (see Fig. 5) where the band at the K point is singly degenerate when the SOC effect is not included. After considering the SOC effect, it becomes double degenerate. To see how the bands split after considering the SOC effect, we should consider the decomposition of the direct-product representation between a one-dimensional representation ($\Gamma_1$ or $\Gamma_2$) and spin representation $D_{1/2}$. Thus, we should get the character of the spin representation $D_{1/2}$ first. If an operation O is a proper rotation with the rotation angle $\alpha$, the character of spin representation $D_{1/2}$ can be obtained from the formula:

$$\chi^j(\alpha)=\frac{\sin{(j+\frac{1}{2})\alpha}}{\sin{(\frac{\alpha}{2})}}$$

Here, j is chosen to be 1/2. For improper rotation, one can decompose it to the product between a proper rotation and inversion I. Since the $D_{1/2}$ spin representation is invariant under the inversion I [54], we also can obtain the character of spin representation $D_{1/2}$ by using the above formula. The obtained characters of spin representation $D_{1/2}$ are shown in Table I. One can easily get $\Gamma_1 \otimes D_{1/2} = \Gamma_6$ and $\Gamma_2 \otimes D_{1/2} = \Gamma_6$. This result indicates that after considering the SOC effect, the singly degenerated band will become double degenerate bands at the K point. Near the K(K') point, the dispersion is linear. This can be explained from group theory. First, after considering SOC effect, the bands belong to the $\Gamma_6$ representation at the K point. Since the Kronecker product $\Gamma_6 \otimes \Gamma_6$ can be decomposed into $\Gamma_1 + \Gamma_2 + \Gamma_3$ and the IR $\Gamma_3$ is a vector representation of the little group $C_{3v}$, indicating that the linear dispersion must be guaranteed. Thus, there is a band crossing point at the K (K') point after the SOC effect is included for a singly degenerate band before the SOC effect is included.

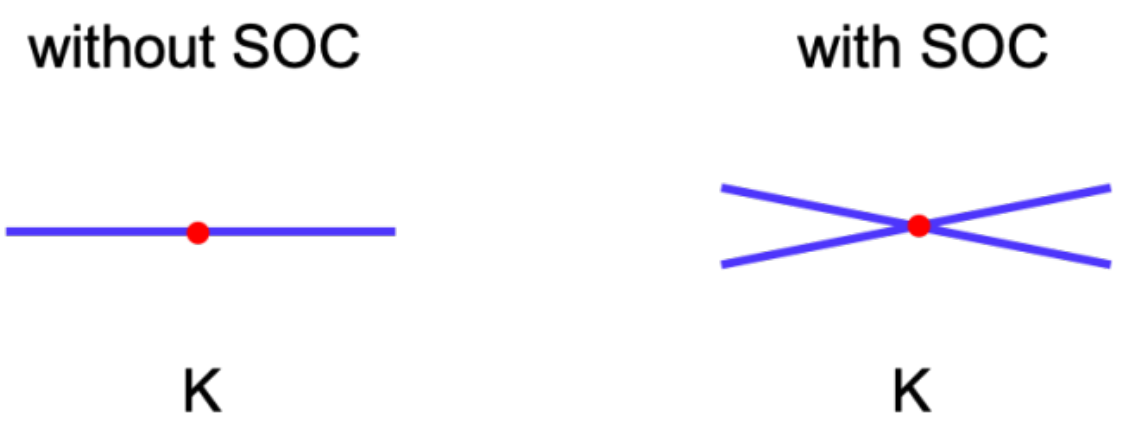

FIG. 5 Without SOC effect, for the K (K') point, the band is singly degenerate. After considering the SOC effect, the band is double degenerate at the K point and split away from the K point, leading to BCPs. This kind of BCPs requires the same filling condition (odd number of electrons) as that of time-reversal symmetry protected BCPs.

Now, we turn to the second case where the bands at the K point belong to a two-dimensional representation. The single point group representation of $C_{3v}$ only has one two-fold IR $\Gamma_3$. The direct product representation between $\Gamma_3$ and $D_{1/2}$ is $\Gamma_3 \otimes D_{1/2} = \Gamma_4 + \Gamma_5 + \Gamma_6$. This indicates that the band will split into a double degenerate band and two single degenerate bands after considering the SOC effect. Near the K(K') point, the dispersion is linear, which can be proved as follows. Considering the SOC effect, we have $\Gamma_3 \otimes D_{1/2} = \Gamma_4 + \Gamma_5 + \Gamma_6$. The two crossing bands belongs to the two-fold IR $\Gamma_6$. Since the Kronecker product $\Gamma_6 \otimes \Gamma_6$ can be decomposed into $\Gamma_1 + \Gamma_2 + \Gamma_3$ and the $\Gamma_3$ IR is a vector representation of the little group $C_{3v}$, indicating that the linear dispersion must be guaranteed. Thus, there is a band crossing point at the K (K') point after the SOC effect is included for double degenerate bands before the SOC effect is included.

|  | $O_1$ | $O_2$ | $O_3$ | $O_4$ | $O_5$ | $O_6$ |  |
|---|---|---|---|---|---|---|---|
| $\Gamma_1$ | 1 | 1 | 1 | 1 | 1 | 1 | $O_1$: E |
| $\Gamma_2$ | 1 | 1 | -1 | 1 | 1 | -1 | $O_2$: $C_{3(001)}^1, \overline{C_{3(001)}^2}$ |
| $\Gamma_3$ | 2 | -1 | 0 | 2 | -1 | 0 | $O_3$: $m_{1-10}, m_{120}, m_{210}$ |
| $\Gamma_4$ | 1 | -1 | -i | -1 | 1 | i | $O_4$: $\overline{E}$ |
| $\Gamma_5$ | 1 | -1 | i | -1 | 1 | -i | $O_5$: $C_{3(001)}^2, \overline{C_{3(001)}^1}$ |
| $\Gamma_6$ | 2 | 1 | 0 | -2 | -1 | 0 |  |
| $D_{1/2}$ | 2 | 1 | 0 | -2 | -1 | 0 | $O_6$: $\overline{m_{1-10}}, \overline{m_{120}}, \overline{m_{210}}$ |

Table I The character table for the $C_{3v}$ double point group.

## APPENDIX B: Details of first-principles calculations

In this work, density functional theory (DFT) method is used for structural relaxation and electronic structure calculation. The ion-electron interaction is treated by the projector augmented-wave (PAW)[55] technique as implemented in the Vienna ab initio simulation package[56]. The exchange-correlation potential is treated by the generalized gradient approximation (GGA) in the Perdew-Burke-Ernzerhof (PBE) form [57]. For structural relaxation, all the atoms are allowed to relax until atomic forces are smaller than 0.01 eV/Å. The 2D k-mesh is generated by the Monkhorst-Pack scheme. To avoid the interaction between neighboring layers, the vacuum thickness is chosen to be 13 Å.

## APPENDIX C: Model Hamiltonian derived with the invariant method

In this part, we derive the effective model Hamiltonian with the invariant method. According to group theory, applying an operator $\hat{g} \in G$ ($G$ is the symmetry group of the system) to the basis functions $\{\varphi_i\}$, the new functions can be expressed as a linear combination of the basis functions:

$$\hat{g}\varphi_i = \sum_j D_{ji}(\hat{g})\varphi_j$$

These matrices $D(\hat{g})$ form a representation of $G$.

Based on the theory of invariants[58], Hamiltonian must satisfy the condition:

$$D^{-1}(\hat{g})H(\kappa)D(\hat{g}) = H(\hat{g}^{-1}\kappa), \hat{g} \in G$$

In general, $D(\hat{g})$ is reducible and can be decomposed into different IRs.

$$D(\hat{g}) = \sum_{i=1}^{n} {}^{\oplus}\Gamma^{\mathrm{i}}(\hat{g})$$

Then, the Hamiltonian can be decomposed into submatrix according to the IRs:

$$\hat{H}(\kappa) = \begin{pmatrix} \hat{H}^{11} & \cdots & \hat{H}^{1n} \\ \vdots & \ddots & \vdots \\ \hat{H}^{n1} & \cdots & \hat{H}^{nn} \end{pmatrix}$$

Since the $D(\hat{g})$ is block-diagonal, invariance of the Hamiltonian yields the following relations for each submatrix:

$$\hat{\Gamma}^{\alpha}(\hat{g}^{-1})\hat{H}^{\alpha\beta}(\kappa)\hat{\Gamma}^{\beta}(\hat{g}) = \hat{H}^{\alpha\beta}(\hat{g}^{-1}\kappa)$$

where $\hat{H}^{\alpha\beta}(\kappa)$ is a $d_\alpha \times d_\beta$ matrix, $d_\alpha$ is the dimension of IR $\alpha$.

The submatrices $\hat{H}^{\alpha\beta}(\kappa)$ can be formed by the products between representation matrices $\hat{X}_l^{\gamma}(\tau|\alpha,\beta)$ and the corresponding basis functions $H_l^{\bar{\gamma}}(\kappa)$, which transform according to complex conjugate IR $\bar{\gamma}$.

$$\hat{H}^{\alpha\beta}(\kappa) = \sum_{\tau}\sum_{\gamma,l} \hat{X}_l^{\gamma}(\tau|\alpha,\beta)H_l^{\bar{\gamma}}(\kappa)$$

$$\hat{\Gamma}^{\bar{\alpha}} \otimes \hat{\Gamma}^{\beta} = \sum_{\gamma} {}^{\oplus} a_\gamma \hat{\Gamma}^{\gamma}$$

$$a_\gamma = \frac{1}{|G|}\sum_{g \in G} \overline{\chi^{\gamma}}(g)\,\overline{\chi^{\alpha}}(g)\chi^{\beta}(g)$$

$$\tau = 1,2, \ldots a_\gamma$$

$X_l^{\gamma}(\tau|\alpha,\beta)$ can be figured out by:

$$\left(X_l^{\gamma}(\tau|\alpha,\beta)\right)_{i,j} \sim \begin{pmatrix} \alpha \\ i \end{pmatrix.\tau \left| \begin{matrix} \bar{\gamma}\beta \\ l\, j \end{matrix} \right)$$

where $\begin{pmatrix} \alpha \\ i \end{pmatrix.\tau \left| \begin{matrix} \bar{\gamma}\beta \\ l\, j \end{matrix} \right)$ is the Clebsch-Gordan coefficient.

At the *K* point, the little point group is $C_{3v}$ and the crossing bands belong to the two-dimensional IR $\Gamma_3$ without considering the SOC effect. We can define two bases as $|\phi_1\rangle$ and $|\phi_2\rangle$, which transform as the two-dimensional IR $\Gamma_3$. Based on these two bases, the representation matrix of space group operators at the K point can be written

as

$$D(g_1)=\begin{pmatrix}1&0\\0&1\end{pmatrix},D(g_2)=\begin{pmatrix}e^{i\theta}&0\\0&e^{-i\theta}\end{pmatrix},D(g_3)=\begin{pmatrix}e^{-i\theta}&0\\0&e^{i\theta}\end{pmatrix}$$

$$D(g_4)=\begin{pmatrix}0&1\\1&0\end{pmatrix},D(g_5)=\begin{pmatrix}0&e^{-i\theta}\\e^{i\theta}&0\end{pmatrix},D(g_6)=\begin{pmatrix}0&e^{i\theta}\\e^{-i\theta}&0\end{pmatrix},\theta=\frac{2\pi}{3}$$

Here, we define $\boldsymbol{a}=\left(\frac{1}{2},-\frac{\sqrt{3}}{2},0\right),\boldsymbol{b}=\left(\frac{1}{2},-\frac{\sqrt{3}}{2},0\right),\boldsymbol{c}=(0,0,1)$. $g_1=E$ is the invariant operator, $g_2=3^{+}_{001}$ is anticlockwise threefold rotation around the $c$ axis, $g_3=3^{-}_{001}$ is clockwise threefold rotation around $\boldsymbol{c}$ axis. $g_4=m_{1\bar{1}0}$ is a reflection perpendicular to vector $\boldsymbol{a}-\boldsymbol{b}$. $g_5=m_{120}$ is reflection perpendicular to vector $\boldsymbol{a}+2\boldsymbol{b}$. $g_6=m_{210}$ is a reflection perpendicular to the vector $2\boldsymbol{a}+\boldsymbol{b}$.

According to the invariant method, $\hat{H}(\kappa)=\hat{H}^{33}$. Decomposing the $\hat{\Gamma}^{\bar{\alpha}}\otimes\hat{\Gamma}^{\beta}$:

$$\bar{\Gamma}_3\otimes\Gamma_3=\Gamma_1\oplus\Gamma_2\oplus\Gamma_3$$

Linear basis functions and constant transform according to $\Gamma_1,\Gamma_2,\Gamma_3$ of $C_{3v}$ are listed in the following table.

| $\gamma$ | $\Gamma_1$ | $\Gamma_2$ | $\Gamma_3$ |
|---|---|---|---|
| $\{H_l^{\bar{\gamma}}(\kappa)\}$ | 1 | | $\{k_+,k_-\}$ |

$$\hat{X}_1^1(1|\Gamma_3,\Gamma_3)\sim I=\begin{pmatrix}1&0\\0&1\end{pmatrix}$$

$$\hat{X}_1^{\Gamma_3}(1|\Gamma_3,\Gamma_3)\sim\begin{pmatrix}\begin{pmatrix}\bar{\Gamma}_3\Gamma_3&\Gamma_3\\1\,1&1\end{pmatrix}&\begin{pmatrix}\bar{\Gamma}_3\Gamma_3&\Gamma_3\\1\,1&2\end{pmatrix}\\\begin{pmatrix}\bar{\Gamma}_3\Gamma_3&\Gamma_3\\1\,2&1\end{pmatrix}&\begin{pmatrix}\bar{\Gamma}_3\Gamma_3&\Gamma_3\\1\,2&2\end{pmatrix}\end{pmatrix}=\begin{pmatrix}0&1\\0&0\end{pmatrix}$$

$$\hat{X}_2^{\Gamma_3}(1|\Gamma_3,\Gamma_3)\sim\begin{pmatrix}\begin{pmatrix}\bar{\Gamma}_3\Gamma_3&\Gamma_3\\2\,1&1\end{pmatrix}&\begin{pmatrix}\bar{\Gamma}_3\Gamma_3&\Gamma_3\\2\,1&2\end{pmatrix}\\\begin{pmatrix}\bar{\Gamma}_3\Gamma_3&\Gamma_3\\2\,2&1\end{pmatrix}&\begin{pmatrix}\bar{\Gamma}_3\Gamma_3&\Gamma_3\\2\,2&2\end{pmatrix}\end{pmatrix}=\begin{pmatrix}0&0\\1&0\end{pmatrix}$$

$$\hat{H}^{33}(\kappa)=\hat{X}_1^{\Gamma_1}(1|\Gamma_3,\Gamma_3)+0+\hat{X}_1^{\Gamma_3}(1|\Gamma_3,\Gamma_3)k_++\hat{X}_2^{\Gamma_3}(1|\Gamma_3,\Gamma_3)k_-$$

$$=\begin{pmatrix}C_0&C_2k_+\\C_2k_-&C_0\end{pmatrix}.$$

It is clear that the band dispersion is linear near the K point when the SOC effect is not included.

Considering the SOC effect, $|\phi_1\rangle$ and $|\phi_2\rangle$ split into $|\phi_{1\uparrow}\rangle, |\phi_{1\downarrow}\rangle$ and $|\phi_{2\uparrow}\rangle, |\phi_{2\downarrow}\rangle$. The corresponding double-valued representation turns into $\Gamma_3 \otimes \Gamma_6 = \Gamma_4 \oplus \Gamma_5 \oplus \Gamma_6$, thus

$$\hat{H}(\kappa) = \begin{pmatrix} \hat{H}^{44} & \hat{H}^{45} & \hat{H}^{46} \\ \hat{H}^{54} & \hat{H}^{55} & \hat{H}^{56} \\ \hat{H}^{64} & \hat{H}^{65} & \hat{H}^{66} \end{pmatrix}$$

Decomposing the $\hat{\Gamma}^{\bar{\alpha}} \otimes \hat{\Gamma}^{\beta}$ for each submatrix:

| | $\Gamma_4$ | $\Gamma_5$ | $\Gamma_6$ |
|---|---|---|---|
| $\bar{\Gamma}_4 = \Gamma_5$ | $\Gamma_1$ | $\Gamma_2$ | $\Gamma_3$ |
| $\bar{\Gamma}_5 = \Gamma_4$ | $\Gamma_2$ | $\Gamma_1$ | $\Gamma_3$ |
| $\bar{\Gamma}_6 = \Gamma_6$ | $\Gamma_3$ | $\Gamma_3$ | $\Gamma_1 \oplus \Gamma_2 \oplus \Gamma_3$ |

According to the invariant method,

$$\hat{H}^{44} = X_1^{\Gamma_1}(1|\Gamma_4, \Gamma_4)$$

$$\hat{H}^{55} = X_1^{\Gamma_1}(1|\Gamma_5, \Gamma_5)$$

$$\hat{H}^{45} = 0$$

$$\hat{H}^{46} = X_1^{\Gamma_3}(1|\Gamma_4, \Gamma_6)k_+ + X_2^{\Gamma_3}(1|\Gamma_4, \Gamma_6)k_-$$

$$\hat{H}^{56} = X_1^{\Gamma_3}(1|\Gamma_5, \Gamma_6)k_+ + X_2^{\Gamma_3}(1|\Gamma_5, \Gamma_6)k_-$$

$$\hat{H}^{66} = X_1^{\Gamma_1}(1|\Gamma_6, \Gamma_6) + 0 + X_1^{\Gamma_3}(1|\Gamma_6, \Gamma_6)k_+ + X_2^{\Gamma_3}(1|\Gamma_6, \Gamma_6)k_-$$

The remain submatrices can be constructed by the hermiticity of Hamiltonian

$$X_1^{\Gamma_3}(1|\Gamma_4, \Gamma_6) = \left( \begin{pmatrix} \Gamma_4 & 1 \\ 1 & \end{pmatrix} \middle| \begin{matrix} \Gamma_3 \Gamma_6 \\ 1\ 1 \end{matrix} \right) \quad \left( \begin{pmatrix} \Gamma_4 & 1 \\ 1 & \end{pmatrix} \middle| \begin{matrix} \Gamma_3 \Gamma_6 \\ 1\ 2 \end{matrix} \right) = \begin{pmatrix} 0 & \frac{1}{\sqrt{2}} \end{pmatrix}$$

$$X_2^{\Gamma_3}(1|\Gamma_4, \Gamma_6) = \left( \begin{pmatrix} \Gamma_4 & 1 \\ 1 & \end{pmatrix} \middle| \begin{matrix} \Gamma_3 \Gamma_6 \\ 2\ 1 \end{matrix} \right) \quad \left( \begin{pmatrix} \Gamma_4 & 1 \\ 1 & \end{pmatrix} \middle| \begin{matrix} \Gamma_3 \Gamma_6 \\ 2\ 2 \end{matrix} \right) = \begin{pmatrix} \frac{i}{\sqrt{2}} & 0 \end{pmatrix}$$

$$X_1^{\Gamma_3}(1|\Gamma_5, \Gamma_6) = \left( \begin{pmatrix} \Gamma_5 & 1 \\ 1 & \end{pmatrix} \middle| \begin{matrix} \Gamma_3 \Gamma_6 \\ 1\ 1 \end{matrix} \right) \quad \left( \begin{pmatrix} \Gamma_5 & 1 \\ 1 & \end{pmatrix} \middle| \begin{matrix} \Gamma_3 \Gamma_6 \\ 1\ 2 \end{matrix} \right) = \begin{pmatrix} 0 & \frac{1}{\sqrt{2}} \end{pmatrix}$$

$$X_2^{\Gamma_3}(1|\Gamma_5, \Gamma_6) = \left( \begin{pmatrix} \Gamma_5 & 1 \\ 1 & \end{pmatrix} \middle| \begin{matrix} \Gamma_3 \Gamma_6 \\ 2\ 1 \end{matrix} \right) \quad \left( \begin{pmatrix} \Gamma_5 & 1 \\ 1 & \end{pmatrix} \middle| \begin{matrix} \Gamma_3 \Gamma_6 \\ 2\ 2 \end{matrix} \right) = \begin{pmatrix} \frac{-i}{\sqrt{2}} & 0 \end{pmatrix}$$

$$X_1^{\Gamma_3}(1|\Gamma_6, \Gamma_6) = \begin{pmatrix} \left( \begin{matrix} \Gamma_6 & 1 \\ 1 & \end{matrix} \middle| \begin{matrix} \Gamma_3 \Gamma_6 \\ 1\ 1 \end{matrix} \right) & \left( \begin{matrix} \Gamma_6 & 1 \\ 1 & \end{matrix} \middle| \begin{matrix} \Gamma_3 \Gamma_6 \\ 1\ 2 \end{matrix} \right) \\ \left( \begin{matrix} \Gamma_6 & 1 \\ 2 & \end{matrix} \middle| \begin{matrix} \Gamma_3 \Gamma_6 \\ 1\ 1 \end{matrix} \right) & \left( \begin{matrix} \Gamma_6 & 1 \\ 2 & \end{matrix} \middle| \begin{matrix} \Gamma_3 \Gamma_6 \\ 1\ 2 \end{matrix} \right) \end{pmatrix} = \begin{pmatrix} 0 & 0 \\ -1 & 0 \end{pmatrix}$$

$$X_2^{\Gamma_3}(1|\Gamma_6,\Gamma_6)=\begin{pmatrix}\left(\begin{array}{c|c}\Gamma_6 1 & \Gamma_3\Gamma_6\\ 1 & 2\;1\end{array}\right) & \left(\begin{array}{c|c}\Gamma_6 1 & \Gamma_3\Gamma_6\\ 1 & 2\;2\end{array}\right)\\ \left(\begin{array}{c|c}\Gamma_6 1 & \Gamma_3\Gamma_6\\ 2 & 2\;1\end{array}\right) & \left(\begin{array}{c|c}\Gamma_6 1 & \Gamma_3\Gamma_6\\ 2 & 2\;2\end{array}\right)\end{pmatrix}=\begin{pmatrix}0 & 1\\ 0 & 0\end{pmatrix}$$

Take these matrices together,

$$\hat{H}(\kappa)=\begin{pmatrix} a_0 & 0 & ia_1k_- & a_1k_+\\ 0 & b_0 & -ib_1k_- & b_1k_+\\ -ia_1k_+ & ib_1k_+ & c_0 & ic_1k_-\\ a_1k_- & b_1k_- & -ic_1k_+ & c_0\end{pmatrix}$$

.

Note that the DFT calculations show that the two bands near the Fermi level belong to the $\Gamma_6$ representation. The Dirac-like form of $\hat{H}^{66}$ suggests that the dispersion of the two bands near the Fermi level is linear around the K-point.

Note that the bases $(|\psi_1\rangle,|\psi_2\rangle,|\psi_3\rangle,|\psi_4\rangle)$ of the above Hamiltonian transform according to $\Gamma_4\oplus\Gamma_5\oplus\Gamma_6$, which are different from the bases $|\phi_{1\uparrow}\rangle,|\phi_{1\downarrow}\rangle$, $|\phi_{2\uparrow}\rangle$, and $|\phi_{2\downarrow}\rangle$, Now, we transform them to $|\phi_{1\uparrow}\rangle,|\phi_{1\downarrow}\rangle$ and $|\phi_{2\uparrow}\rangle,|\phi_{2\downarrow}\rangle$, that is transforms according to $\Gamma_3\otimes\Gamma_6$. Suppose that

$$(|\phi_{1\uparrow}\rangle,|\phi_{1\downarrow}\rangle,|\phi_{2\uparrow}\rangle,|\phi_{2\downarrow}\rangle)=(|\psi_1\rangle,|\psi_2\rangle,|\psi_3\rangle,|\psi_4\rangle)U$$

$U$ can be obtained by using the Clebsch-Gordan coefficient,

$$U=\begin{pmatrix}\left(\begin{array}{c|c}\Gamma_3\Gamma_6 & \Gamma_4 1\\ 1\;1 & 1\end{array}\right) & \left(\begin{array}{c|c}\Gamma_3\Gamma_6 & \Gamma_5 1\\ 1\;1 & 1\end{array}\right) & \left(\begin{array}{c|c}\Gamma_3\Gamma_6 & \Gamma_6 1\\ 1\;1 & 1\end{array}\right) & \left(\begin{array}{c|c}\Gamma_3\Gamma_6 & \Gamma_6 1\\ 1\;1 & 2\end{array}\right)\\ \left(\begin{array}{c|c}\Gamma_3\Gamma_6 & \Gamma_4 1\\ 1\;2 & 1\end{array}\right) & \left(\begin{array}{c|c}\Gamma_3\Gamma_6 & \Gamma_5 1\\ 1\;2 & 1\end{array}\right) & \left(\begin{array}{c|c}\Gamma_3\Gamma_6 & \Gamma_6 1\\ 1\;2 & 1\end{array}\right) & \left(\begin{array}{c|c}\Gamma_3\Gamma_6 & \Gamma_6 1\\ 1\;2 & 2\end{array}\right)\\ \left(\begin{array}{c|c}\Gamma_3\Gamma_6 & \Gamma_4 1\\ 2\;1 & 1\end{array}\right) & \left(\begin{array}{c|c}\Gamma_3\Gamma_6 & \Gamma_5 1\\ 2\;1 & 1\end{array}\right) & \left(\begin{array}{c|c}\Gamma_3\Gamma_6 & \Gamma_6 1\\ 2\;1 & 1\end{array}\right) & \left(\begin{array}{c|c}\Gamma_3\Gamma_6 & \Gamma_6 1\\ 2\;1 & 2\end{array}\right)\\ \left(\begin{array}{c|c}\Gamma_3\Gamma_6 & \Gamma_4 1\\ 2\;2 & 1\end{array}\right) & \left(\begin{array}{c|c}\Gamma_3\Gamma_6 & \Gamma_5 1\\ 2\;2 & 1\end{array}\right) & \left(\begin{array}{c|c}\Gamma_3\Gamma_6 & \Gamma_6 1\\ 2\;2 & 1\end{array}\right) & \left(\begin{array}{c|c}\Gamma_3\Gamma_6 & \Gamma_6 1\\ 2\;2 & 2\end{array}\right)\end{pmatrix}^{\dagger}$$

$$=\begin{pmatrix}0 & 1/\sqrt{2} & i/\sqrt{2} & 0\\ 0 & 1/\sqrt{2} & -i/\sqrt{2} & 0\\ 0 & 0 & 0 & 1\\ -1 & 0 & 0 & 0\end{pmatrix}$$

Hence, Hamiltonian with basis $|\phi_{1\uparrow}\rangle,|\phi_{1\downarrow}\rangle,|\phi_{2\uparrow}\rangle,|\phi_{2\downarrow}\rangle$ is

$$\hat{H}'(\kappa)=U^{\dagger}\hat{H}(\kappa)U$$

$$
= \begin{pmatrix} c_0 & -\frac{a_1+b_1}{\sqrt{2}}k_- & -i\frac{a_1-b_1}{\sqrt{2}}k_- & ic_1k_+ \\ -\frac{a_1+b_1}{\sqrt{2}}k_+ & \frac{a_0+b_0}{2} & i\frac{a_0-b_0}{2} & i\frac{a_1-b_1}{\sqrt{2}}k_- \\ i\frac{a_1-b_1}{\sqrt{2}}k_+ & -i\frac{a_0-b_0}{2} & \frac{a_0+b_0}{2} & \frac{a_1+b_1}{\sqrt{2}}k_- \\ -ic_1k_- & -i\frac{a_1-b_1}{\sqrt{2}}k_+ & \frac{a_1+b_1}{\sqrt{2}}k_+ & c_0 \end{pmatrix}
$$

One can see that the dominant term is the first order of $k_+$ and $k_-$, thus, the dispersion is linear, in agreement with the above group theory analysis.

## APPENDIX D: Edge states of zigzag and armchair $FeB_2$ nanoribbons

The edge states calculated without SOC effect for the zigzag and armchair edges are shown in Fig. 6. For zigzag edge, the two BCPs in the bulk system are projected onto different points, which are connected by non-trivial edge states (Fig. 6a) that can be regarded as nearly-free-electron (NFE) Shockley surface states[59]. In addition, these edge states are relatively long in *k* space due to the large distance between the K and K' points in the bulk system. However, for the armchair edge, since the two BCPs in the bulk system are projected onto the same point (Fig. 6b), the non-trivial edge states cannot be visualized here (strictly speaking, the edge states shrink to a point). We note that edge states of $FeB_2$ ribbons are similar to that (Fig. 7) of graphene ribbons [60]. After considering the SOC effect, the edge states of $FeB_2$ split, possibly due to the Rashba effect (Fig. 6c and 6d). We also calculate the Berry phase by using a close loop encircling a band crossing point in the 2D $FeB_2$ system, and find that the result is π, indicating that the band crossing point is topological non-trivial.

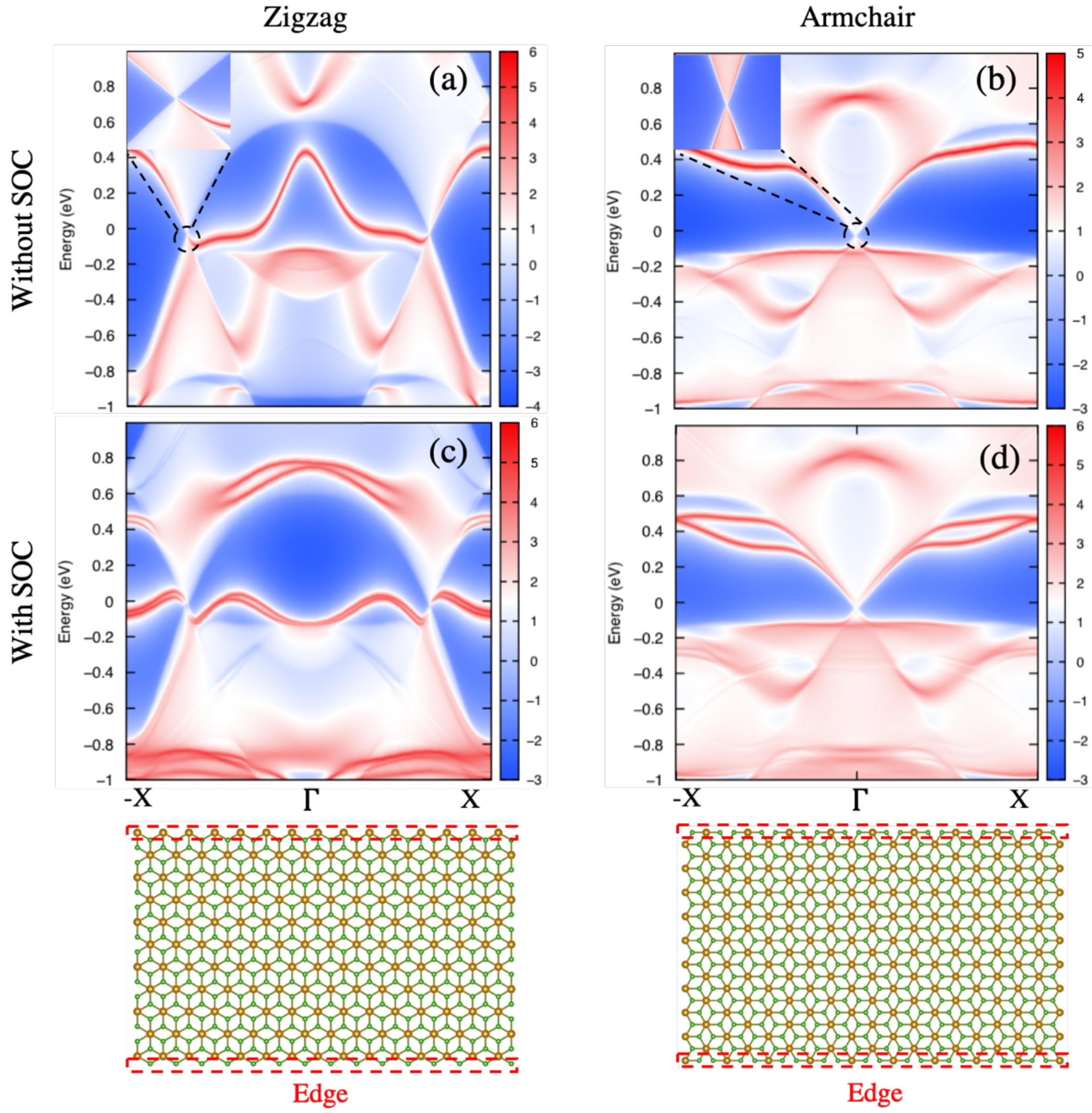

FIG. 6 Edge states of armchair and zigzag $FeB_2$ ribbons. The insert in (a) indicate that there exists an edge state which connects two different BCPs. This is non-trivial edge state. For armchair, the edge state shrinks to a point at Γ point. The other edge states in (b) are trivial. (c) and (d) are edge states with the SOC effect taken into account.

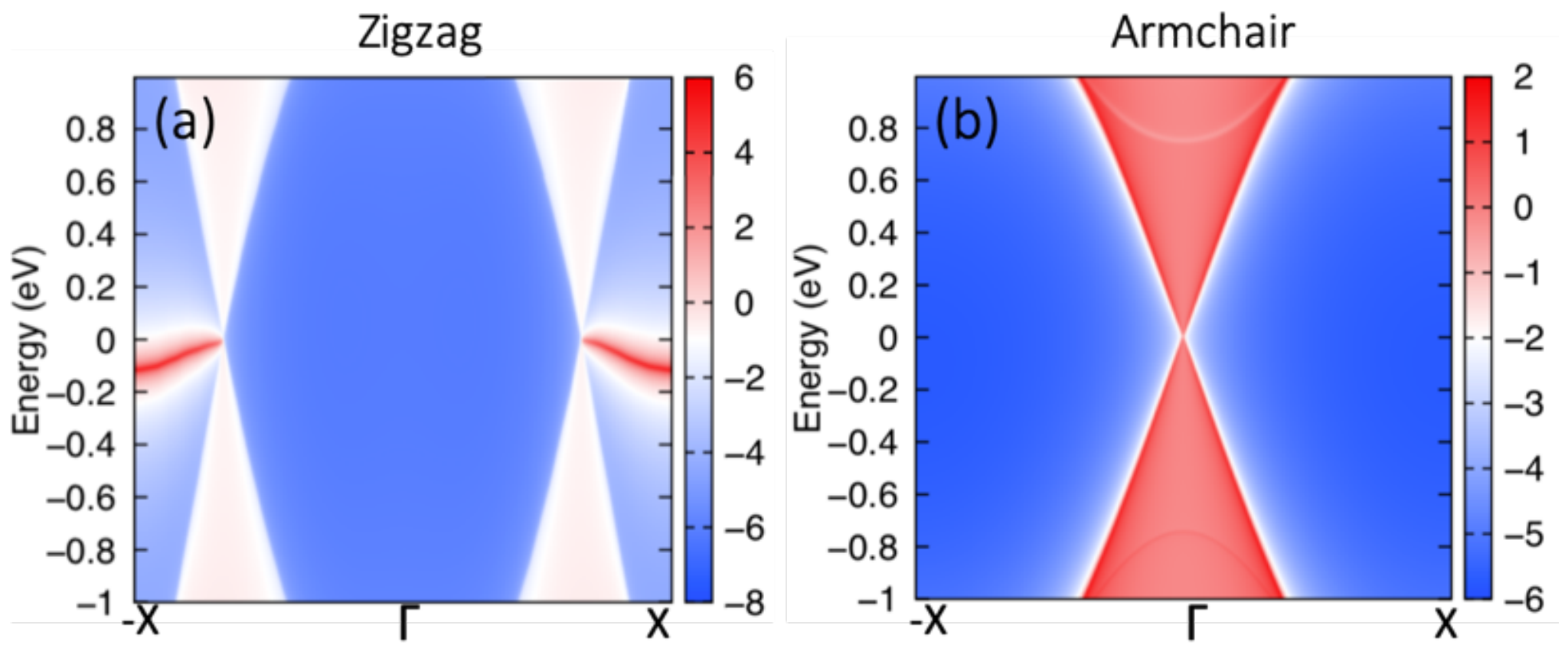

FIG. 7 Edge states of zigzag (a) and armchair (b) graphene ribbons without considering the SOC effect. For the armchair edge, the two Dirac points in the bulk system project onto the same point, thus, there is only one Dirac point for the armchair ribbon.